\nonstopmode
\documentclass[aip,pop,reprint,superscriptaddress]{revtex4-1}

\usepackage{graphicx}
\usepackage{epstopdf}
\usepackage{natbib}
\usepackage{amsmath,amssymb,amsfonts}
\usepackage[usenames,dvipsnames]{color}
\usepackage{subcaption}
\usepackage[normalem]{ulem}

\usepackage{tikz} 
\usetikzlibrary{arrows.meta, positioning}
 
\begin{document}

\title{Hybrid PIC-fluid model for numerical simulation of laser-plasma interaction}

\author{Andrey Sladkov}
\email{andrey.sladkov1992@gmail.com}
\affiliation{DUSOFT LAB LLC, Russia, Dzerzhinsk, Chapaev street 22, 606007}
\begin{abstract}

A hybrid PIC–fluid model is proposed for three dimensional numerical simulation of laser–plasma interaction. Ions are treated kinetically, electrons as a ten-moment fluid, capturing ion-scale dynamics, pressure anisotropy, and non-Maxwellian distributions efficiently. A laser-envelope model handles energy deposition and ponderomotive heating without resolving optical oscillations. Collisional and ionisation processes ensure self-consistent evolution of energy and charge states. The model is implemented in the AKAM code, providing a scalable framework that bridges fully kinetic and fluid approaches for high-energy-density and laboratory plasma applications.

\end{abstract}

\maketitle

\section{Introduction} \label{sec:intro}

Hybrid Particle--In--Cell (PIC) -- fluid modeling approaches, which treat ions kinetically and electrons as a charge-neutralizing fluid, have become the only practical pathway for performing three dimensional, end-to-end, full-scale plasma simulations in regimes where both ion-kinetic physics and large-scale electromagnetic evolution matter. Fully kinetic PIC approaches, while physically complete, are computationally prohibitive for most laboratory and space-plasma scenarios. Resolving electron plasma and cyclotron frequencies, Debye-length charge separation, and realistic electron-to-ion mass ratios requires extreme temporal and spatial resolution, making system-size simulations infeasible even on modern architectures~\cite{birdsall2018}. Fully fluid approaches, such as magnetohydrodynamics (MHD), achieve tractability only through restrictive assumptions—typically Maxwellian closures, isotropic pressure, and single-fluid formulations. These simplifications suppress key processes, including finite-Larmor-radius effects, anisotropy-driven instabilities, and wave–particle interactions, particularly when pressure transport is non-negligible~\cite{chew1956}.

Hybrid modeling addresses this challenge by preserving ion phase-space structure and non-Maxwellian distributions while allowing extended electron closures, including tensor pressure evolution~\cite{sladkov2021}, which becomes important when electron anisotropy or nongyrotropy is dynamically relevant. This capability is crucial for accurately modeling processes such as magnetic sheath and shock formation, mirror and firehose instabilities, temperature-anisotropy-driven wave generation, and ion heating in expanding or fast-flowing plasmas~\cite{winske2003}. It is equally critical for modern high-energy-density and extreme ultraviolet (XUV)-source environments~\cite{versolato2022}, where rapidly expanding, strongly magnetised plasmas develop pronounced ion and electron anisotropy that can hardly be correctly treated within MHD yet remain inaccessible to full-PIC scales. Consequently, hybrid PIC methods uniquely deliver the physical fidelity and computational scalability required for predictive simulations across contemporary laboratory, astrophysical, and XUV-source plasma applications.

The earliest hybrid codes were developed to model high-altitude nuclear explosions~\cite{mcnamara1968}, where expanding plasma clouds interacted with the geomagnetic field, requiring a simultaneous description of large-scale electromagnetic fields and small-scale ion motion. Subsequent pioneering studies ~\cite{forslund1971,chodura1974,chodura1975} established hybrid simulations as a robust tool for investigating laminar collisionless shocks, ion reflection, and magnetic-field steepening in high-Mach-number flows. Beyond space physics, hybrid approaches have also been successfully applied to laboratory high-energy-density plasmas, including pinch implosions and magnetic-field diffusion experiments~\cite{sgro1976}, demonstrating their versatility for both experimental and theoretical studies.

Hybrid codes strike a balance between physical fidelity and computational efficiency, making them suitable for comprehensive simulations of large-scale plasma systems. The hybrid approach has been successfully applied to astrophysical debris–plasma interactions~\cite{winske2007}, the study of shock formation criteria~\cite{winske2012}, and laboratory experiments employing laser-driven magnetic pistons to generate quasi-perpendicular shocks~\cite{larson2015, schaeffer2014}. Recent work~\cite{keenan2022} highlights the rich interplay of electromagnetic waves, kinetic instabilities, and anisotropy-driven ion heating that emerges when fast plasma flows interact with a magnetised background. Classic numerical studies of plasma cloud expansion in the external magnetic field~\cite{golubev1979,bashurin1984} further underscore the importance of magnetic pressure gradients and ion reflection in shaping the expanding plasmas. Beyond these traditional applications, hybrid simulations now provide a practical framework for modeling modern high-energy-density and XUV-source experiments, enabling predictive simulations across multiple scales.

In summary, hybrid PIC -- fluid model offers a versatile and computationally efficient framework for investigating laser–plasma interactions across a wide range of regimes. By treating ions kinetically while modeling electrons as a fluid, these codes capture the essential ion-scale dynamics and particle-level effects, while remaining tractable for full-scale, end-to-end simulations. Hybrid PIC simulations thus provide both predictive capability and physical insight, bridging the gap between theory, numerical modeling, and experimental observation in contemporary laser–plasma research.

\section{Numerical model} \label{sec:model}

The hybrid code AKAM~\cite{akamGitHub} builds upon established principles~\cite{winske2003} used in previous codes such as HECKLE~\cite{smets2011} and AKA~\cite{sladkov2020}. The code AKA~\cite{sladkov2020} has been successfully employed in a variety of laboratory plasma studies, including investigations of magnetised plasma jets~\cite{zemskov2024, zemskov2026}, interacting plasma flows~\cite{korzhimanov2025}, magnetic reconnection~\cite{bolanos2022}, magnetic field compression~\cite{sladkov2024}, and accretion-channel modeling in young stellar objects~\cite{burdonov2022}. AKAM~\cite{akamGitHub} builds on this established framework by incorporating additional physics relevant to laser–plasma experiments, including explicit laser beam deposition, electron–ion collisional processes, and dynamic ionisation. These enhancements allow AKAM to capture the coupled evolution of laser-driven plasmas, electromagnetic fields, and ionisation states in high-energy-density regimes, enabling end-to-end simulations of modern laboratory experiments.

It keeps the ion description at the particle level and considers electrons as a neutralising fluid whose density equals that of the ions ($n_e=n_i=n$). Electrons are described by the ten-moment model. Those ten moments are density - $n$, bulk velocity - $\mathbf V_e$ and the six-component pressure tensor - $\mathbf P_e$. The electromagnetic fields are treated in the low-frequency (Darwin) approximation, neglecting the displacement current, an electron Ohm's law~:

\begin{equation}
\mathbf E = -\mathbf V_i \times \mathbf B + \frac{1}{e n}(\mathbf J \times \mathbf B - \boldsymbol{\nabla} . \, \mathbf P_e) + \eta \mathbf J \label{ohm}
\end{equation}

In Eq.~(\ref{ohm}), $\mathbf V_i$ is the ion bulk velocity, $n$ is the electron density, $\mathbf J$ is the total current density equal to the curl of $\mathbf B$, electron inertia is neglected. For the six-component pressure tensor evolution equation, the explicit subcycling integration scheme is used~\cite{sladkov2021}. Electromagnetic fields are calculated on two staggered grids using a predictor-corrector scheme~\cite{winske1986}. The dynamics of the ions is solved using a first-order interpolation scheme~\cite{boris1972}.

In the numerical model, the magnetic field and the density are normalised to $B_0$ and $n_0$ respectively, times are normalised to the inverse of ion gyrofrequency $\Omega_0^{-1}$ (calculated using the magnetic field $B_0$), lengths are normalised to the ion inertial length $d_0$ (calculated using the density $n_0$), and velocities are normalised to the Alfv\'en velocity $V_0$ (calculated using $B_0$ and $n_0$). Mass and charge are normalised to the ion ones. The normalisation of the other quantities follows from these ones.

In summary, the hybrid numerical model provides a self-consistent framework to capture the essential physics of ion kinetics. The chosen normalisation, boundary conditions, and parameterisation can ensure consistency with the experimental setups, allowing direct comparison between numerical predictions and laboratory observations.

\section{Laser-plasma coupling}
\label{sec:env}
Supposing in laser–plasma interaction problems of our interest the optical frequency is much faster than any hydrodynamic or plasma-response timescale. This scale separation allows the laser field to be written in terms of a rapidly oscillating carrier multiplied by a slowly varying complex envelope. The vector potential is represented as
\begin{equation}
  \hat{A}(\mathbf{x},t)
  =
  \Re\!\left[
    \tilde{A}(\mathbf{x},t)\,
    e^{\,i(k_0 x - \omega_0 t)}
  \right],
  \label{eq:env_ansatz}
\end{equation}
where $\tilde{A}(\mathbf{x},t)$ is the laser complex envelope, $k_0=\omega_0/c$, and $\omega_0$ is the central laser frequency. The envelope $\tilde{A}$ varies slowly in space and time relative to the oscillatory carrier.

Starting from the D’Alembert equation for the laser field, and substituting the ansatz \eqref{eq:env_ansatz},  retaining only slowly varying terms and discarding the rapidly oscillatory components. This leads to the envelope equation
\begin{equation}
  \nabla^2 \tilde{A}
  + 2i\!\left(\partial_x\tilde{A} + \partial_t\tilde{A}\right)
  - \partial_t^2 \tilde{A}
  = \chi(\mathbf{x},t)\,\tilde{A}
  \label{eq:env_main}
\end{equation}
where $\chi(\mathbf{x},t)$ is the plasma susceptibility,
\begin{equation}
  \chi(\mathbf{x},t)
  = -\frac{n_e(\mathbf{x},t)}{n_{\mathrm{cr}}}, \qquad
  n_{\mathrm{cr}}
  = \frac{m_e \varepsilon_0 \omega_0^2}{e^2}
  \label{eq:env_chi}
\end{equation}
The susceptibility produces refraction and self-focusing of the laser in the plasma, while diffraction enters through the transverse Laplacian in \eqref{eq:env_main}. The term $2i(\partial_x+\partial_t)\tilde{A}$ encodes group-velocity propagation of the envelope. The equation is normalised separately where $c=1$.

For hydrodynamic plasma evolution, the laser couples to the plasma primarily through the ponderomotive force, which depends on the cycle-averaged intensity. To compute this force efficiently within the envelope framework the cycle-averaged electric field amplitude is used. Starting from the envelope representation, the corresponding electric field is
\begin{equation}
  \mathbf{E}(\mathbf{x},t)
  = -\partial_t \hat{A}
  \label{eq:env_efield_exact}
\end{equation}

Equation~\eqref{eq:env_main} is advanced on a uniform Cartesian grid using a finite-difference time-domain (FDTD) scheme. The laser envelope $\tilde{A}(\mathbf{x},t)$ is discretised at cell-centered grid points $\mathbf{x}_{i,j,k}$ and advanced explicitly in time.

\subsection{Temporal Discretization}

Time integration is performed using a second-order centered finite-difference scheme. Denoting $\tilde{A}^n_{i,j,k}$ as the envelope at time $t^n = n\Delta t$, the second-order time derivative is approximated by
\begin{equation}
  \partial_t^2 \tilde{A}
  \;\approx\;
  \frac{\tilde{A}^{n+1}_{i,j,k}
        -2\tilde{A}^{n}_{i,j,k}
        +\tilde{A}^{n-1}_{i,j,k}}{\Delta t^2},
  \label{eq:env_time2}
\end{equation}
while the first-order time derivative appearing in the group-velocity term is treated using a centered difference,
\begin{equation}
  \partial_t \tilde{A}
  \;\approx\;
  \frac{\tilde{A}^{n}_{i,j,k}-\tilde{A}^{n-1}_{i,j,k}}{\Delta t}.
  \label{eq:env_time1}
\end{equation}

\subsection{Spatial Discretisation}

The spatial derivatives are computed on a uniform Cartesian grid with spacings $(\Delta x,\Delta y,\Delta z)$. The Laplacian operator is discretised using second-order finite differences in all spatial directions. Along the laser propagation direction $x$, a stabilised fourth-order stencil~\cite{massimo2019} is employed to reduce numerical dispersion. The discrete Laplacian reads
\begin{align}
  \nabla^2 \tilde{A}
  &\approx
  (1+\delta)
  \frac{\tilde{A}_{i+1,j,k}^n
       -2\tilde{A}_{i,j,k}^n
       +\tilde{A}_{i-1,j,k}^n}{\Delta x^2}
  \nonumber \\
  &\quad
  -\frac{\delta}{4}
  \frac{\tilde{A}_{i+2,j,k}^n
       -2\tilde{A}_{i,j,k}^n
       +\tilde{A}_{i-2,j,k}^n}{\Delta x^2}
  \nonumber \\
  &\quad
  +\frac{\tilde{A}_{i,j+1,k}^n
       -2\tilde{A}_{i,j,k}^n
       +\tilde{A}_{i,j-1,k}^n}{\Delta y^2}
  \nonumber \\
  &\quad
  +\frac{\tilde{A}_{i,j,k+1}^n
       -2\tilde{A}_{i,j,k}^n
       +\tilde{A}_{i,j,k-1}^n}{\Delta z^2}
  \label{eq:env_laplacian_disc}
\end{align}
where the stabilization parameter
\begin{equation}
  \delta = \frac{1 - (\Delta t/\Delta x)^2}{3}
\end{equation}
controls numerical anisotropy and improves dispersion properties along the
propagation direction.

The longitudinal gradient term is discretized consistently with the stabilized
stencil,
\begin{align}
  \partial_x \tilde{A}
  &\approx
  (1+\delta)\,
  \frac{\tilde{A}_{i+1,j,k}^n-\tilde{A}_{i-1,j,k}^n}{2\Delta x}
  -\frac{\delta}{2}\,
  \frac{\tilde{A}_{i+2,j,k}^n-\tilde{A}_{i-2,j,k}^n}{2\Delta x}.
  \label{eq:env_grad_disc}
\end{align}

\subsection{Fully Discrete Update Equation}

Combining the above discretizations, the envelope equation
\eqref{eq:env_main} is advanced explicitly according to
\begin{align}
  \tilde{A}^{n+1}_{i,j,k}
  &=
  \frac{1+i k_0 \Delta t}
       {1+k_0^2\Delta t^2}
  \Big[
     2\tilde{A}^{n}_{i,j,k}
     - (1+i k_0 \Delta t)\tilde{A}^{n-1}_{i,j,k}
  \nonumber \\
  &\qquad
     + \Delta t^2
       \big(
          \nabla^2 \tilde{A}^{n}_{i,j,k}
          +2i k_0\,\partial_x \tilde{A}^{n}_{i,j,k}
          -\chi^{n}_{i,j,k}\tilde{A}^{n}_{i,j,k}
       \big)
  \Big],
  \label{eq:env_update}
\end{align}
where $\chi^{n}_{i,j,k} = -n_e^{n}/n_{\mathrm{cr}}$ is evaluated explicitly from
the plasma density at the current time step.

A predictor--corrector scheme is employed to achieve second-order accuracy in time. The predictor stage extrapolates the envelope forward, while the corrector averages predicted and previous values to suppress numerical noise.

\subsection{Electric Field Reconstruction}

The physical laser electric field is formally obtained from the vector potential, within the envelope formulation, a discrete approximation of this relation can be written as
\begin{equation}
  E_{\mathrm{laser}}^{n}
  =
  \Re\!\left[
    \frac{\tilde{A}^{n}-\tilde{A}^{n-1}}{\Delta t}
    - i\omega_0 \tilde{A}^{n}
  \right],
  \label{eq:env_efield_disc}
\end{equation}
which combines a finite-difference evaluation of the slow temporal variation of the envelope with the dominant contribution arising from the fast carrier oscillation.

In practice, plasma coupling mechanisms such as the ponderomotive force and inverse bremsstrahlung depend on the electric-field intensity averaged over an optical period. Owing to the strong scale separation $|\partial_t \tilde{A}| \ll \omega_0 |\tilde{A}|$, the cycle-averaged electric
field satisfies
\begin{equation}
  \langle |E_{\mathrm{laser}}|^2 \rangle
  \approx
  \frac{\omega_0^2}{2}\,|\tilde{A}|^2,
  \label{eq:env_efield_avg}
\end{equation}
which is independent of the carrier phase. Consequently, all laser--plasma energy and momentum exchange terms are evaluated using the envelope magnitude rather than the instantaneous oscillatory electric field which could be used for the ions push step.

\subsection{Inverse Bremsstrahlung absorption}

The envelope couples to the plasma through the susceptibility $\chi$ and
through collisional (inverse Bremsstrahlung) absorption.  
The local heating rate is
\begin{equation}
  Q_{\mathrm{IB}}
  =
  \frac{n_e \nu_{ei}}{m_e}
  \frac{ \langle |E_{\mathrm{laser}}|^2 \rangle}{\omega_0^2 + \nu_{ei}^2},
  \label{eq:env_ib}
\end{equation}
where $\nu_{ei}$ is the electron–ion collision frequency,
\begin{equation}
  \nu_{ei}
  \sim
  \frac{
     Z^2 n_i \ln\Lambda
  }
  {(m_i T_e)^{3/2}}\label{eifreq}
\end{equation}
where $Z$ - charge, $\ln\Lambda$ - Coulomb logarithm. This absorbed energy is added to the electron thermal energy and contributes to the evolution of the electron pressure, closing the coupling between the laser beam and electron hydrodynamics.
\\
\\
\\
The laser envelope model described above captures diffraction, refraction, density-gradient effects, and collisional absorption while remaining computationally efficient. The scheme is explicit, second-order accurate in space and time, and naturally suited for domain-decomposed parallel simulation of laser–plasma interaction.

\section{Electron ion collision operator}\label{sec:eicoll}

The interaction between ion particles and the fluid electron background is treated through a collision operator that self-consistently exchanges thermal energy and momentum between ions and electrons. This approach combines a Landau--Spitzer--type relaxation model for electron--ion temperature equilibration. For each ion species $s$ in each cell, the pre-collision scalar ion temperature $T_{i,s}^{(0)}$ is computed from particle velocities relative to the cell-centered ion bulk velocity $\mathbf{u}_s$. The electron pressure tensor from the fluid solver is reduced to a scalar electron temperature $T_e$, with a density-dependent edge-profile factor applied to prevent unphysical heating in regions of very low electron density.

The electron--ion collision frequency (Eq.~\ref{eifreq})
defines the associated electron--ion relaxation time $\tau_{ie} = 1/\nu_{ei}$. The target ion temperature after one PIC timestep $\Delta t$ is obtained from the exact solution of the relaxation equation:
\begin{equation}
  T_{i,s}^{\mathrm{tgt}}
  = T_e - \left(T_e - T_{i,s}^{(0)}\right) \exp\Big(-\frac{\Delta t}{\tau_{ie}}\Big),
\end{equation}
with a small-temperature limiter applied to very cold electrons to avoid excessive energy transfer.

Particle velocities are then rescaled to match the target ion temperature while preserving the bulk velocity:
\begin{equation}
  \mathbf{v}_{i,p}^{\mathrm{new}} = \mathbf{u}_s + \lambda \left(\mathbf{v}_{i,p}^{\mathrm{old}} - \mathbf{u}_s \right),
  \qquad
  \lambda = \sqrt{\frac{T_{i,s}^{\mathrm{tgt}}}{T_{i,s}^{\mathrm{curr}}}},
\end{equation}
where $T_{i,s}^{\mathrm{curr}}$ is the particle-based temperature of species $s$ in the cell. For cells with multiple particles, stochastic pairwise scattering is applied in a Takizuka--Abe fashion~\cite{takizuka1977}, scaled by the fractional timestep $\Delta t/\tau_{ie}$. This ensures isotropic redistribution of thermal energy while conserving momentum.

Finally, the change in ion thermal pressure due to collisions is projected back onto the fluid grid to modify the electron pressure tensor, completing the self-consistent coupling between ion particles and the fluid electron background. The collision operator conserves total energy to numerical accuracy. This approach accurately models collisional energy exchange, temperature relaxation, and pressure feedback in hybrid PIC--fluid simulations of plasma dynamics.

\section{Ionisation Processes}
\label{sec:ionisation}

The ionisation state of each particle is evolved based on the local electron
conditions, namely density $n_e$ and pressure $p_e$. Each particle's charge $Z_p$
is updated during a timestep $\Delta t$ according to the net effect of ionisation
and recombination processes:
\begin{equation}
  Z_p^{\mathrm{new}} = Z_p + \Delta Z, \qquad
  Z_\mathrm{min} \le Z_p^{\mathrm{new}} \le Z_\mathrm{max},
\end{equation}
where $\Delta Z$ represents the net change in ionisation state during $\Delta t$.

The ionisation increment $\Delta Z$ can be computed using generic rate laws, for
example
\begin{equation}
  \Delta Z = (\Gamma_\mathrm{ion} - \Gamma_\mathrm{rec})\,\Delta t,
\end{equation}
where $\Gamma_\mathrm{ion}$ and $\Gamma_\mathrm{rec}$ are the ionisation and recombination rates, respectively. These rates may depend on the local electron temperature $T_e$, electron density $n_e$, and particle charge $Z_p$, and different models (e.g., collisional, tunnel, or field ionisation) can be employed depending on the plasma regime of interest.

In practice, a stochastic rounding procedure can be applied to ensure that $\Delta Z$ remains an integer change in charge, while maintaining the charge within physical limits. This allows for a computationally efficient, yet physically consistent treatment of ionisation in a PIC framework, without constraining the choice of specific ionisation and recombination laws.

By coupling the particle charges to the local plasma conditions, this model enables self-consistent evolution of the plasma ionisation state and energy exchange with electrons.

\section{Overall Simulation Algorithm}
\label{sec:algorithm}

The main solver advances the plasma--laser system through time by performing a sequence of coupled operations on particles, fields, and hydrodynamic quantities. At each timestep, the algorithm follows a structured loop that ensures self-consistent evolution of particle motion, electromagnetic fields, and plasma state.

\subsection*{Algorithm Loop}

The main steps performed in a single timestep are:

\begin{enumerate}
    \item \textbf{Ion collisions:} if collisional effects are enabled, ion-ion collisions are computed to relax ion velocities toward local
          equilibrium.
    \item \textbf{Particle push:} all particles are advanced in phase space under the action of electromagnetic fields.
    \item \textbf{Ionisation:} particle charges are updated according to the local plasma conditions.
    \item \textbf{Moment gathering:} particle quantities (density, momentum) are accumulated onto the computational grid.
    \item \textbf{Electron-ion collisions:} electrons interact with ions, updating thermal energies and momentum exchange.
    \item \textbf{Laser interaction:} laser fields are deposited and envelope  propagation is computed; particles may receive additional momentum
          from laser acceleration.
    \item \textbf{Electron pressure pumping:} the effect of collisional (inverse Bremsstrahlung) absorption.
    \item \textbf{Electromagnetic field update:} magnetic and electric fields are advanced using Maxwell's equations in a leapfrog scheme.
    \item \textbf{Closure relations:} plasma pressure and other derived quantities are updated based on moments and particle states.
\end{enumerate}

This loop is repeated for each timestep, ensuring the self-consistent evolution of particles, fields, and plasma properties. Conditional operations (e.g., collisions or laser deposition) are only executed when relevant for the simulation setup.

\section{Test case}
\begin{figure*}[ht!]
  \centering
  \includegraphics[width=0.99\textwidth]{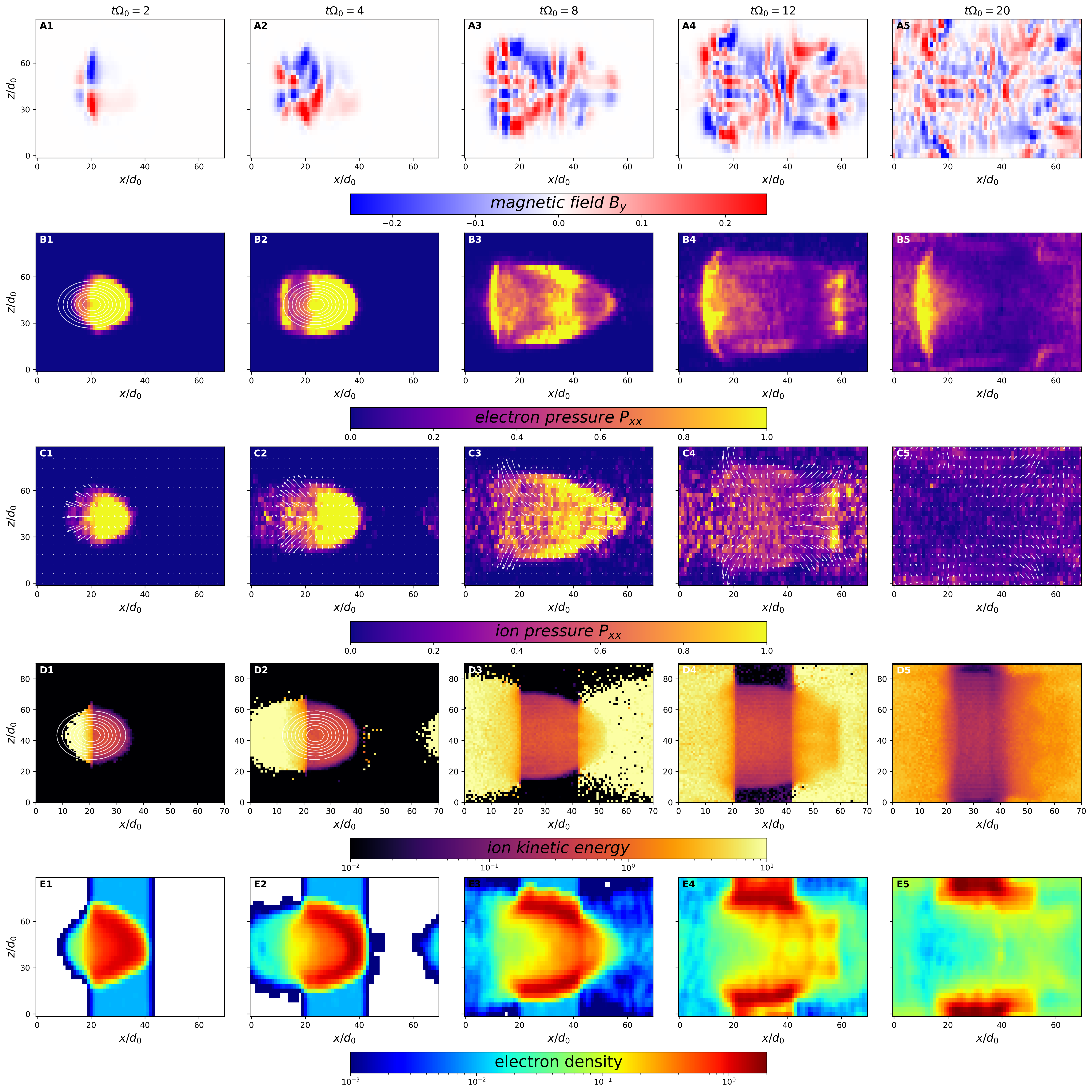}
  \caption{ Spatiotemporal evolution of fields and particle energy in the laser–plasma interaction test case at consecutive output times (columns from left to right).
  \textbf{(A)} Out-of-plane component of magnetic-field $B_y$ in the central $y$ slice.
  \textbf{(B)} Electron pressure tensor component $P_{xx}$. White lines represent laser beam potential isocontours.
  \textbf{(C)} Ion pressure tensor component $P_{xx}$. Arrows represent ion flow velocity.
  \textbf{(D)} Particle kinetic energy density integrated in $y$ direction, computed from particle velocities.  \textbf{(E)} Electron density.}
  \label{fig:test_case_fields_energy}
\end{figure*}

The algorithm was tested using a laser slab configuration; the simulation ran on a dual-core laptop for approximately seven minutes. The simulation domain is 3D rectangular box with sizes $L_X = 70$, $L_Y = 90$, $L_Z = 90$ and periodic boundary conditions in all directions. A grid of 70$\times$30$\times$30 points corresponds to a mesh size of 1$d_0$ along the x direction and 3$d_0$ along the y and z. The time step was set to $5\times10^{-2}$. Each cell representing the target fraction contains 100 particles. The background plasma was frozen, with one particle per cell. The plasma consists of protons with mass 1 and maximal charge 1. The simulation models the laser ablation process and the associated ionisation dynamics, starting from an initially low plasma charge of $10^{-2}$. Total simulation time is 20$\Omega_0^{-1}$. In the pressure tensor evolution equation~\cite{sladkov2021}, there are artificially increased electron mass, zero divergence of the electron heat flux and a modest smoothing to stabilise noise.

The laser pulse is injected along the $x$ direction and described using the envelope formulation introduced in Section~\ref{sec:env}. The temporal profile is Gaussian, with a full width at half maximum of $10\,d_0$, laser frequency $\omega_0$ = 10$\Omega_0$, and the peak normalised vector potential is $a_0 = 1$. The laser is focused near the center of the domain at $x = L_X/2$, resulting in a converging beam propagating through the initially cold target plasma. After $t\Omega_0 = 7.5$ the potential is set to zero. Critical density in (Eq.~\ref{eq:env_update}) is 100$n_0$, minimal potential threshold for the electron-beam interaction is $10^{-2}$. The electron-ion energy transfer operator uses uniform collision frequency $\tau_{ie}\Omega_0$ = 1.

To assess the self-consistent evolution of fields and particle energisation, a set of diagnostics is analysed at five representative times, shown in Fig.~\ref{fig:test_case_fields_energy}. The first row displays the transverse magnetic-field component $B_y$ in the central $y$ slice. As the laser propagates and deposits energy, localised magnetic structures emerge and grow in amplitude. The clear Biermann-battery~\cite{biermann1951} magnetic field is seen near the target edge at $t\Omega_0 = 2.5$ which then expands and becomes fragmented due to anisotropy-driven plasma instabilities.

The second and third rows show the electron and ion pressure tensor components $P_{xx}$, respectively. At early times, the electron pressure exhibits a compact structure localized near the laser focal region. As the interaction proceeds, the electron pressure broadens and propagates forward, indicating efficient momentum transfer from the laser field to the electron population. Ion heating displays a more diffuse spatial distribution, consistent with indirect energy transfer mediated by electron--ion collisions and electrostatic fields. Notably, the pressure maps reveal the formation of three distinct shock fronts: parallel and antiparallel to the laser trace, and cylindrically symmetric inside the target traversing tangentially to the target surface. These shocks are particularly evident in both the electron and ion pressure evolution, highlighting the multistage energy deposition and momentum transfer from electrons to ions.

The fourth row presents the particle kinetic energy density, computed by binning particle energies in the $(x,z)$ plane. The resulting maps demonstrate a progressive increase in particle energy and spatial spreading along the laser propagation direction. The fifth row shows the electron density distribution, revealing the plasma expansion and a shock front moving along laser propagation direction at $t\Omega_0 = 4$. Early in the interaction, the density profile remains compact, while at later times it develops a characteristic multi-lobed structure. The density evolution, along with pressure and energy diagnostics, clearly illustrates the coupled dynamics of laser-driven compression, shock formation, and plasma expansion.

Overall, the correlated evolution of magnetic fields, pressure, particle energy, and electron density confirms that the envelope-based laser model accurately captures the dominant mechanisms of laser absorption, shock generation, and momentum transfer. This particular simulation is a synthetic run designed to demonstrate the consistency of the scheme, with box sizes chosen to be close enough to examine how plasmas interact through periodic boundaries and unphysical frozen-in background. The emergence of multiple shocks and the associated density modulations further validates the envelope approach as an efficient and robust framework for modeling laser ablation in high-energy-density plasma conditions.

\section{Discussion}\label{sec:diss}

\begin{figure}[htbp]
    \centering
    \includegraphics[width=0.48\textwidth]{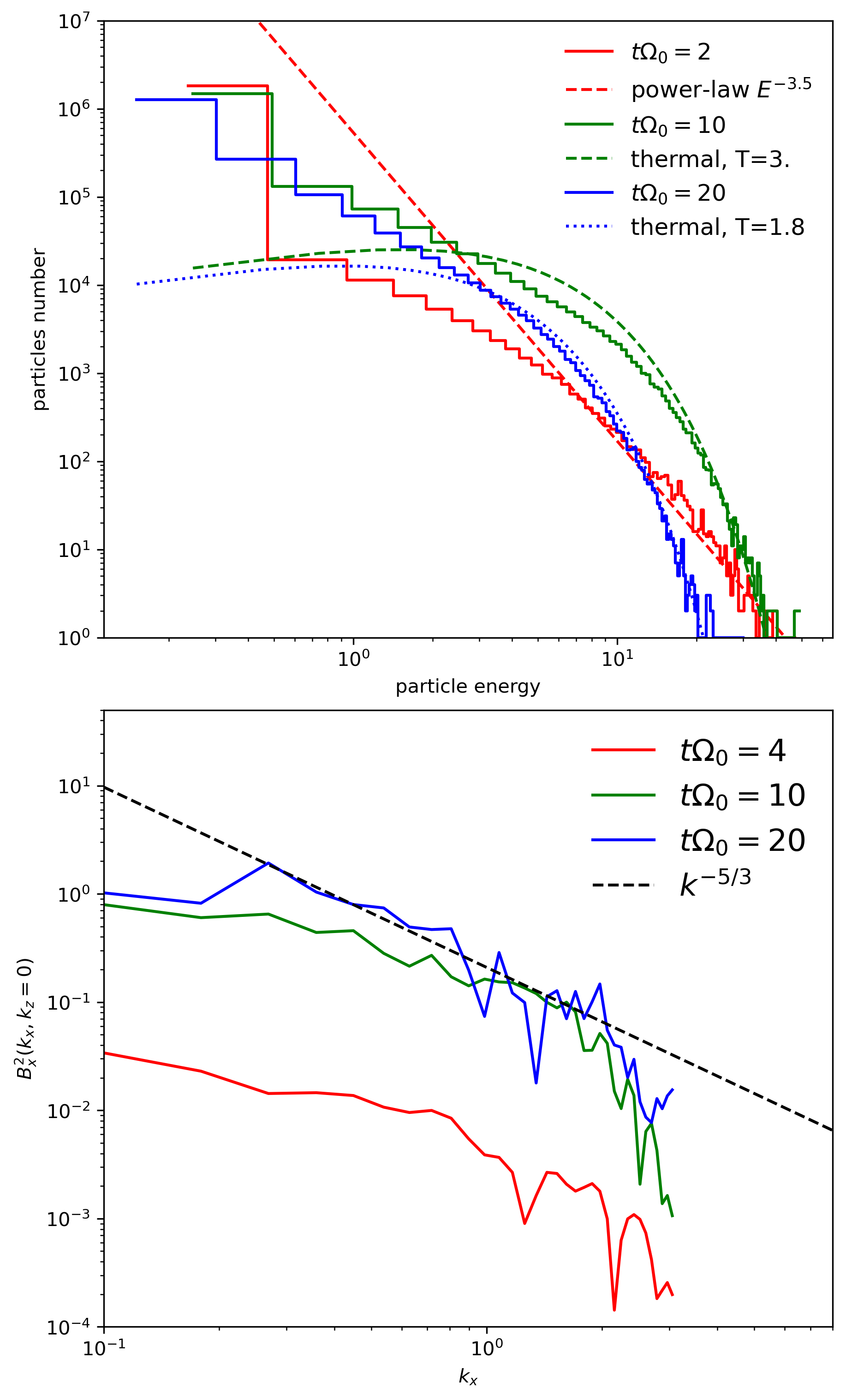}
    \caption{
    Upper panel: Ions energy distributions at different times $t\Omega_0 = 2, 10, 20$ with corresponding power-law and thermal fits. 
    Lower panel: 1D slice of a 2D Fourier spectrum of the square of magnetic field $B^2_x(k_x, k_z=0)$ at the central plane as on Fig.~\ref{fig:test_case_fields_energy}.
    }
    \label{fig:energy_fft}
\end{figure}

The test case demonstrates that the AKAM hybrid PIC--fluid framework accurately captures key ion-scale kinetics and laser--plasma interactions while remaining computationally efficient. Fig.~\ref{fig:energy_fft} upper panel shows ion energy distribution, at early times spectrum can be approximated by the power law, later it becomes thermal with cutoff shifting to the low energies. Electron heating occurs rapidly and is initially localised near the laser focal region, whereas ions experience delayed and more diffuse energisation through electron--ion collisions and electrostatic coupling, producing non-Maxwellian energy tails consistent with indirect acceleration mechanisms.

The simulations reveal the emergence of localised magnetic structures, consistent with current filamentation driven by electron pressure anisotropies. Fig.~\ref{fig:energy_fft} lower panel shows a slice of a 2d Fourier spectrum for square of magnetic field $B^2_x$, at the end of simulation the long-wave part becomes closer to the canonical cascade regime $\sim k^{-5/3}$ showing how the hybrid approach captures the transition from sub-ion to multiple ion scales. This evolution, together with the correlated pressure and energy features, demonstrates that the hybrid PIC--fluid framework faithfully captures both the macroscopic plasma response and the underlying kinetic dynamics.

The envelope-based laser model plays a crucial role in this dynamics, efficiently mediating energy deposition, ponderomotive heating while avoiding spurious carrier-scale oscillations and allowing larger timesteps. Coupled ionisation and inverse Bremsstrahlung operators ensure a self-consistent evolution of charge states and electron thermal energy, which is subsequently transferred to ions through collisional relaxation. The resulting correlations between magnetic structures, pressure anisotropies, electron density, and ion energization confirm the self-consistent interplay of fields, particles, and plasma state. The results confirms that hybrid PIC--fluid simulations with envelope laser coupling can provide a robust and predictive framework for modeling high-energy-density and laser-ablation plasmas.

\section{Conclusions}
\label{sec:conclusion}

AKAM is a 3D hybrid PIC–fluid framework for modeling laser–plasma interactions in high-energy-density regimes. By treating ions kinetically and electrons as a ten-moment fluid, the code captures essential ion-scale dynamics, pressure anisotropies, and non-Maxwellian distributions while remaining computationally efficient. The envelope-based laser model resolves energy deposition, ponderomotive heating without resolving optical cycles, while coupled collisional and ionisation processes provide self-consistent energy exchange and dynamic charge-state evolution.

3D simulations demonstrate self-generated magnetic-structure formation, rapid electron heating, delayed ion energisation, and high-energy ion tails. These capabilities make AKAM particularly well suited for applications involving XUV sources, laser-driven plasma jets, and high-energy-density laboratory experiments, where accurate modeling of expanding, strongly magnetised, and anisotropic plasmas is critical. By bridging the gap between fully kinetic and fluid approaches, AKAM offers a predictive and scalable tool for both experimental design and fundamental studies in extreme plasma environments.

\bibliographystyle{alpha}
\bibliography{biblio.bib}

\end{document}